\documentstyle[aps,pre,multicol]{revtex}
\input epsf
\begin{document}
\draft
\title{Analysis of a dissipative model of self-organized criticality
with random neighbors}
\author{Marie-Line Chabanol and Vincent Hakim}
\address{ Laboratoire de Physique Statistique\dag, ENS,
 24 rue Lhomond, 75231 Paris Cedex 05, France.\\
\dag associ\'e au CNRS et aux Universit\'es Paris VI et VII}
\date{\today}
\maketitle
\begin{abstract}
We analyze a random neighbor version of the OFC stick-slip model.
We find that the mean avalanche
size is finite as soon as dissipation exists in the bulk but that
this size grows exponentially fast when dissipation tends to zero.
\end{abstract}


\begin{multicols}{2}
It is an appealing idea that many power laws observed in nature arise
from an intrinsic trend of a large class of extended non-equilibrium systems 
to evolve toward critical points \cite{btw}. This
concept of self-organized criticality (SOC)  has therefore
attracted much interest 
and
implicit assumptions of the original model 
have been subjected to intense 
scrutiny.
Several early SOC explanations assigned a crucial role to
strict bulk conservation \cite{obh,carl}. Non-conserving models
which show SOC behavior 
have since been found \cite{dros,bs} but in several cases the effect
of a small dissipation remains unclear.
An interesting model where dissipation is controlled by
a parameter $\alpha$, referred hereafter as the 
OFC model,
has been introduced in \cite{ofc} as a simplified version of previous
modelling of fault dynamics \cite{bur}.
Numerical results and supporting arguments
appear to indicate that the OFC model exhibits power-law distributed
avalanches in the dissipative range of $\alpha$ values below the conserving
 $\alpha_0$ \cite{ofc,mt}. 
A random neighbor version of the OFC model has been studied in
\cite{lj} and numerical evidence of SOC behavior has similarly been
found for $\alpha_c<\alpha<\alpha_0$. 
Our aim here is to
analyze this simpler version of the OFC model. 
In contrast to \cite{lj}, we find
that avalanches are of finite size up to the conserving limit
$\alpha=\alpha_0$ but that their mean size grows exponentially fast as
$\alpha \rightarrow \alpha_0$. This may explain our disagreement
with \cite{lj} and can perhaps also serve as a cautionary note about
similar numerical evidence obtained for local lattice models.

 The  model \cite{lj}
consists of a set of $N$ sites, 
to each of which is associated 
an "energy" $x_i$. The dynamics alternates between two phases:\\
- the loading phase is supposed to take place on a long time scale in the
fault dynamics context. In this phase, all the $x_i$  are below a
threshold and increase continuously and simultaneously
with time. This regime lasts until one energy reaches the
threshold energy which we choose equal to one. At this point, an
avalanche starts and the dynamics enters the avalanche phase.\\
- the avalanche phase is thought to be instantaneous on the time scale
of the loading which can thus be 
neglected. The dynamics is entirely
 governed by energy transfers between different sites. 
For each
$x_i\ge 1$, $K$ different sites $j^{(i)}$ are randomly chosen.
On each one, the energy is increased from 
 $x_{j^{(i)}}$ to $x_{j^{(i)}} + \alpha x_i$ and
then $x_i$ is set to $0$. The process is repeated if some of the new 
energies are above one. When all the site 
energies are smaller than one, the avalanche ends and 
the system returns to the loading phase.\\
The parameter $0\le\alpha \le 1/K$ controls the dissipation
during an avalanche. 
If $\alpha = 1/K$, energy is conserved and the total energy of the
system is constant
during an avalanche. On the contrary, it decreases  for $\alpha<1/K$.

\begin{figure}[htb]
\narrowtext
\centerline{\epsfxsize=90mm \epsfysize=70mm\epsfbox{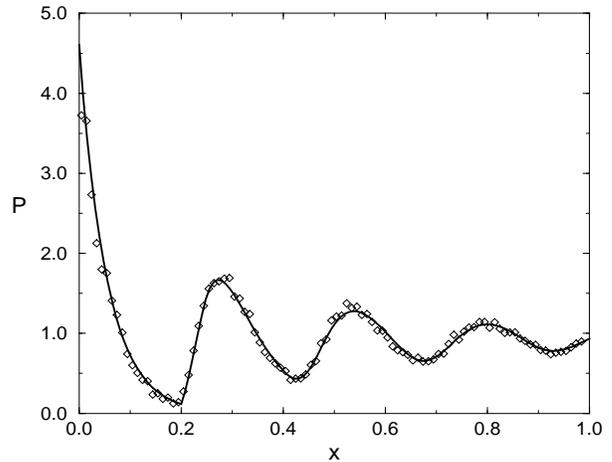}}
\caption
{Stationary distribution obtained by simulation (circles)
 and by solving the equations (\ref{eqa}) and (\ref{eqps}) (continuous line)
for $K=4$ and $\alpha = 0.2$. Simulations results  are the average
of $2\ 10^4$ avalanches on a lattice of $N=5\  10^3$ sites
and the bin size is 0.01.}
\label{theosim}
\end{figure}
Results of simulations \cite{lj} as those reported in Fig.1,
have shown that the probability distribution $P_t(x)$ of site
energies tends at large times toward a non-trivial stationary distribution
$P(x)$.
It is, in fact, possible to obtain the exact evolution equation
obeyed by $P_t(x)$ in the limit $N\rightarrow\infty$ as we now show. We define
the size of an avalanche as the number of topplings during its evolution
(the number of sites where the energy becomes greater or equal to the
threshold).
We suppose that the parameter $\alpha$ is small enough so that 
the mean avalanche size has a finite value when the system size tends to 
infinity.
The two regimes of the dynamics contribute to the evolution of $P_t$
and are considered in turn.
For definiteness, we fix to unity the growth rate due to the constant loading.
 So, in a
time interval $\Delta t$ between $t$ and $t+\Delta t$, the site energies
increase by $\Delta x = \Delta t$. This gives
rise to $M$ avalanches with $M=N P_t(1) \Delta x$ ($N P_t(1)$ is the density
of sites which have their energy just below $1$).
The sites which are
updated during the course of these avalanches
 belong to three distinct classes:\\
-i) the $M$ starting sites of the avalanches the energy of which is 
set to zero,\\
-ii) the sites which through energy redistribution have their 
energies first increased above one and then
set to zero, the total number of which we define to be $M \bar{A}_t$,\\
-iii) those which have their energies increased below one, the final energies
of which are distributed according to a density $M B_t(x)$.\\
Note that we assume that the system is large enough so that
 the probability that a given site has been updated more than once is
negligible.
At $t+\Delta t$, the probability distribution of site energies has
become $P_{t+\Delta t}(x)$ with
\begin{eqnarray}
N P_{t+\Delta t}(x)&=& N P_t(x-\Delta x) - M \delta(x-1) + M B_t(x) 
\nonumber\\
&+&
 M (\bar{A}_t+1)\delta(x) - K M (\bar{A}_t+1) P_t(x).
\end{eqnarray}
The last term on the r.h.s accounts for the fact that
the sites of classes ii) and iii) are picked at random among the $N$ sites
and that their total number is $K M (\bar{A}_t +1)$ (since the energy 
of each site
above threshold is redistributed to $K$ other sites). Taking the limit
$\Delta t \rightarrow 0$, we obtain the evolution equation for $P_t(x)
,\, 0\le x\le 1$ :
\begin{equation}
\!\partial_t P_t(x) +\partial_x P_t(x) = P_t(1)\left[ B_t(x)- K (
\bar{A}_t\!+\!1) P_t(x)\right]
\label{pt}
\end{equation}
together with the injection condition at $x=0$,
\begin{equation}
P_t(0)=P_t(1)\, (\bar{A}_t+1)
\label{ic}
\end{equation}

In order to obtain a closed equation for $P_t(x)$, it remains to determine
the characteristics of the avalanches, $\bar{A}_t$ and $B_t(x)$, in terms
of $P_t(x)$. To this end, we analyze the course of an avalanche in a more
detailed way.
At the n-th step of an avalanche, for each site in class ii) which is set to 
$0$, K sites are randomly chosen. Those which have their energies temporarily
increased
above one belong to class ii) and we denote by
$a_n (x),\, x\ge1$,  
the distribution of their energies above
threshold. 
The avalanche ends when $a_n(x)=0$.
Similarly, we call $b_n(x),\, \alpha\le x<1 $ the distribution of sites of class iii)
produced at the n-th step. This gives therefore for $x\ge 1$, 
\begin{equation}
a_n(x) 
	   = K\int_1^{\frac{1}{1-\alpha}}
 a_{n-1}(y)\, P(x-\alpha y)\,dy 
\label{eqan}
\end{equation}
For $x<1$, one obtains an equation with
the same r.h.s. but with $b_n(x)$ instead on the l.h.s.. In 
Eq.~(\ref{eqan}), the integral upper bound has been taken to be $1/(1-\alpha)$
since a brief study of the series defined by $u_0 = 1$ and $u_{n+1} =
1 + \alpha u_n $ shows that $a_n(x)$ is zero if 
$x\ge 1/(1-\alpha)$.
To compute the evolution of $P_t(x)$, we can restrict ourselves to consider
$B_t(x)$ which is 
the total density  $\sum_{n} b_n$ 
of sites of class iii) averaged over the $M$ avalanches occurring between 
$t$ and $t+\Delta t$.
We similarly define $A_t(x)$ as 
the average over these avalanches of $\sum_{n} a_n$. From (\ref{eqan}),
$A_t(x)$ 
is determined from $P_t(x)$ as
the solution of the linear equation for $x\ge 1$ :
\begin{equation}
A_t(x) = K\left[\int_1^{\frac{1}{1-\alpha}} A_t(y) P_t(x-\alpha y)\,dy  
+ P_t(x-\alpha)\right]
\label{eqa}\\
\end{equation}
For $x<1$, the l.h.s. is replaced  by $B_t(x)$,
\begin{equation}
B_t(x) = K\left[\int_1^{\frac{1}{1-\alpha}} A_t(y) P_t(x-\alpha y)\,dy  
+ P_t(x-\alpha)\right]
\label{eqb}\\
\end{equation}
This gives
$B_t$ in terms of $A_t(x)$ and $P_t(x)$. 
Since $\bar A_t\equiv \int dx\, A_t(x)$, the evolution of $P_t(x)$ is
determined by solving Eq.~(\ref{pt}) together with (\ref{eqa}) using the
expression (\ref{eqb})
for $B_t(x)$. Specializing to the steady state time independent
functions, we finally obtain for $0\le x\le1$,
\begin{equation}
\!\frac{P'(x)}{KP(0)} + P(x) = \frac{\int_1^{\frac{1}{1-\alpha}}\!
 {A}(y) P(x\!-\!\alpha y)\,dy + P(x\!-\!
\alpha)}{\bar A + 1}
\label{eqps}
\end{equation}
where the steady state distribution $A(x)$ is determined from $P(x)$ by
Eq.~(\ref{eqa}) (with the time indices dropped). At this stage, several
simple remarks can be made. The size of an avalanche is the total number
of sites in class ii) plus the starting site so the mean
avalanche size $\bar{s}= 1+ \bar{A}$. 
It is useful to note that the injection condition (\ref{ic}) is
directly obtained  by integrating Eq.~(\ref{eqps}) from $x=0$ to $x=1$ and by
using
$\int_0^1 P=1$ and $\int B = K(\bar{A}+1)-\bar{A}$
 (the last
equality follows from the avalanche rule as noted above but it can
also be derived by adding 
Eq.~(\ref{eqa}) and (\ref{eqb}) and
integrating over $x$).
Eq.~(\ref{ic})
gives the alternative expression of $\bar{s}$ as
$\bar{s}= P(0)/P(1)$. 
There are infinite avalanches with
non-zero probability only if $P(0)$ is
infinite or $P(1)$ is zero. Actually, we find below that
both are true. 
Large avalanches lead to large $P(0)$ but also deplete
the distribution of sites energies away from a small number of given energies
with a
depletion length  proportional to $1/P(0)$ . This leads
$P(1)$ to decrease exponentially fast when $P(0)$ increases.

\begin{figure}[htb]
\narrowtext
\centerline{\epsfxsize=90mm \epsfysize=70mm\epsfbox{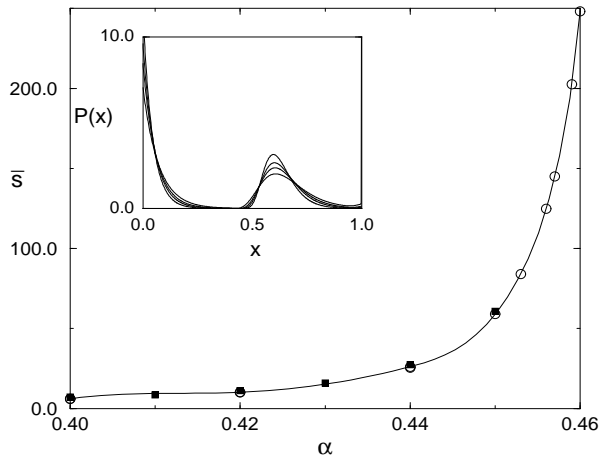}}
\caption
{ Mean avalanche size {\em{vs.}} $\alpha$ for $K=2$ showing
 solutions of
(\ref{eqa}) and (\ref{eqps}) with $\bar{s}=\bar{A}+1$ (circles) and
results of numerical simulations (filled square).
 The line is a guide for the eyes.
Insert : $P(x)$ obtained by solving Eq.~(\ref{eqa},
\ref{eqps}) for $K=2$ and $\alpha=0.45,0.453,0.456, 0.463$.
}
\label{intA}
\end{figure}

\begin{figure}[htb]
\narrowtext
\centerline{\epsfxsize=90mm \epsfysize=70mm\epsfbox{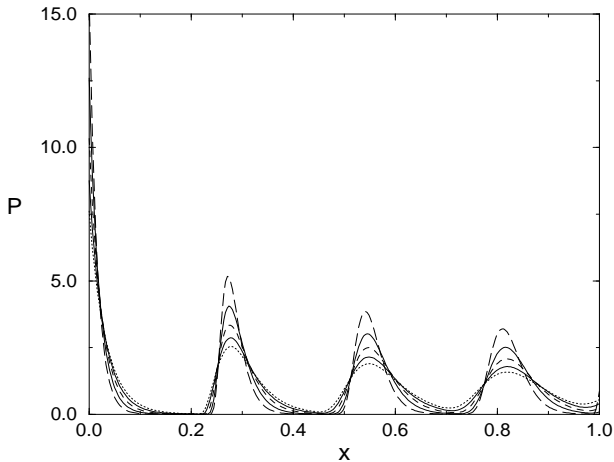}}
\caption
{Stationary distribution obtained by solving Eq.~(\ref{eqa},
\ref{eqps})
for $K=4$ and $\alpha = 0.22, 0.224, 0.228, 0.232,
0.236$.}
\label{theo_r}
\end{figure}

We now turn to the solution of Eq.~(\ref{eqa}) 
and (\ref{eqps}).
First, one can note that the  r.h.s of Eq.~(\ref{eqps}) vanishes for
$0\le x < \alpha$ since $B(x)= 0$ in this range. Therefore, for $x < \alpha$,
one
has the exact form 
$P(x) = P(0) \exp
(-KP(0)x)$ which simply reflects the balance between the
constant site creation at x=0 and the depletion due to site recruitment in
avalanches. Besides this simple result, an analytical determination of
$P(x)$ has eluded us.
We have therefore solved
numerically
Eq.~(\ref{eqa}) and (\ref{eqps}), taking advantage of the known form of
$P(x)$ for $x<\alpha$. Given $A(x)$, this determines the r.h.s of
Eq.~(\ref{eqps}) for $\alpha\le x < 2\alpha$  and thus $P(x)$ in this range.
Continuing the process, the computation of $P(x)$ consists of solving
$K$ linear
inhomogeneous differential equations.
We have thus iterated the following 
process : solve Eq.~(\ref{eqps}) for some $P(0)$ and $A$, normalize $P$ so 
that $\int P = 1$, set $P(0)$ to its new actual value, and use Eq.~(\ref{eqa})
to find the new $A$. The non-trivial satisfaction of relation (\ref{ic})
was used as a check of the computation. 
The computed $P(x)$ for
$K=4$ is shown in
Fig.~\ref{theosim} and agrees
 well with simulations results. For $K=2$, similar
agreement is obtained. 
The computed mean avalanche size agrees with 
the averaged  avalanche size obtained from simulations,  
in the range of $\alpha$ where
they can be
compared, as shown in 
Fig.~\ref{intA} 
for $K=2$.
For 
the largest $\alpha$'s, it was found necessary to
use lattices of $N=2\ 10^4$ and average over $3\ 10^5$ avalanches to ensure 
that convergence to the steady state was reached and that the results were
free from finite size effects.
In both cases, the mean avalanche size grows very quickly when $\alpha$ 
approaches $1/K$. In ref.\cite{lj}, similar results were interpreted as a
divergence of the mean avalanche size at $\alpha_c=0.2255$ for $K=4$. With
our numerical procedure,
the solutions of Eq.~(\ref{eqa}) and (\ref{eqps}) can, however, be
reliably obtained up to $\alpha=0.236$, far above the purported critical
$\alpha_c$. 
The corresponding distributions  $P(x)$ 
are shown in Fig.~\ref{theo_r} for $K=4$ and do not display any singularity
at $\alpha=0.2255$. On the contrary, the four peaks of the distribution
appear to
sharpen smoothly as $\alpha$ increases. This leads us to think that they
smoothly tend to $\delta$ peaks as $\alpha\rightarrow 1/4$.
As shown in the insert of Fig.~\ref{intA}, a similar behavior is observed 
for $K=2$. 

The presence of sharper and sharper peaks puts a heavy demand on
numerical resolution and prevents a direct numerical approach of the limit
$\alpha\rightarrow 1/K$ (with our algorithm, 
at least). To analyze further this limit, we focus for simplicity on
the case $K=2$. For $\alpha=1/2$, $P(x)$ is made of two $\delta$ peaks
located at $x=0$ and $x=1/2$ respectively while $A(x)$ is simply a 
$\delta$ peak at x=1. When $\alpha$ is close to $1/2$, $P(0)$ becomes
large and the derivative term in Eq.~(\ref{eqps}) can balance the other
terms only if $P(x)$ has a fast variation on a scale $1/P(0)$. This is indeed
the case of the exact form of $P(x)$ near $x=0$. We therefore search for
$A$ and $P$ under the form,
\begin{eqnarray}
A(x)&\simeq& \frac{P(0)^2}{P(1)}\, a\left[P(0)(x-1-2 \eta(\alpha))\right]
\label{fass}\\
P(x)&=& P(0)\, \exp(-2P(0) x), \,\, x\le\alpha
\nonumber\\
P(x)&\simeq&
 \frac{1}{2}\,P(0)\,\Pi[P(0)(x-\alpha-\eta
(\alpha))] ,\,\, x\ge\alpha
\label{fpss}
\end{eqnarray}
where $a$ and $\Pi$ are two functions to be determined which have
been normalized so that their integrals equal one and which describe the
broadening for $\alpha\neq 1/2$ of the $\delta$ peaks at $x=1$ and $x=1/2$
respectively. The peak displacement $\eta(\alpha)$ is supposed to tend to
zero as $\alpha\rightarrow 1/2$. Moreover, self-consistency requires that
$P(0) \eta(\alpha)\rightarrow\infty$ when $\alpha$ tends toward $1/2$. This 
allows to neglect the weight of $B(x)/\bar{A}$ 
near $x=1$
(note that $B(x)$ is the continuation of $A(x)$ for
$x<1$) 
as supposed in (\ref{fpss}). 
Substituting the forms (\ref{fass}) and (\ref{fpss}) in Eq.~(\ref{eqa}) and
(\ref{eqps}), we obtain at dominant order,
\begin{eqnarray}
a(x) & = & \int_{-\infty}^{+\infty} \Pi(x + 2C -u/2) a(u)du
\label{eqssa}\\
\Pi(x) &=& 4  e^{-2x} \int_{-\infty}^{2x} a(u) e^u (x - u/2) du 
\label{eqssp}
\end{eqnarray}
where we have defined the constant $C= \lim_{\alpha\rightarrow 1/2} (1/2-\alpha)
P(0)$ and integrated the linear differential equation for $\Pi$. Taking
Fourier transforms of (\ref{eqssa}) and (\ref{eqssp}), one obtains for
$\hat{a}(\omega)=\int dx \exp(i\omega x) a(x)$, the functional equation
\begin{equation}
\hat{a}(\omega)=\frac{\exp(-2 i\omega C)}{(1-i\omega/2)^2}\, \hat{a}^2 
(\omega/2)
\label{eqaf}
\end{equation}
We fix the translational symetry of
Eq.~(\ref{eqssa}) and (\ref{eqssp}) 
[$a(x)\rightarrow a(x+x_0), \Pi(x) \rightarrow \Pi(x+x_0/2)$]
by imposing $\int x a(x) dx =0$. Then, the
unique solution of Eq.~(\ref{eqaf}) without a singularity at $\omega=0$ is
\begin{equation}
\hat{a}(\omega)=\prod_{n=1}^{\infty}\frac{\exp(-2 i\omega C)}{(1-i\omega/2^n)
^{2^n}}
\label{af}
\end{equation}
where convergence of the infinite product enforces $C=1/2$. One can
check that the inverse
Fourier transform $a(x)$ of $\hat{a}(\omega)$ is indeed a real function
and is positive, as it should,
since it is the convolution of the real positive functions
\begin{equation}
v_n(x)= \frac{2^n\left[2^n\left(x+1\right)\right]^{2^n-1}}
{(2^n-1)!}
e^{-2^n\left(x+1\right)}\,\theta(x+1)
\end{equation}
When $x\rightarrow -\infty$, this allows to show
that  $a(x)$ and $\Pi(x)$
tend extremely
quickly towards zero, namely $a(2x)\sim\Pi(x)\sim\exp(- {\mathrm{cst}}
\, 4^{-x})$. Comparing the two estimations (\ref{fpss}) of $P(x)$ near 
$x=\alpha$, one obtains for the peak displacement $P(0)\eta(\alpha)\simeq
|\ln(1/2-\alpha)|/\ln(4)$ as $\alpha\rightarrow 1/2$.
When $x\rightarrow +\infty$, 
the analytic expression (\ref{af}) (or directly
Eq.~(\ref{eqssa}) and (\ref{eqssp})) shows that $a(x)$ and $\Pi(x)$ tend toward
zero as $x\exp(-2x)$. Using this
asymptotic behavior to estimate $P(1)$ gives that 
the mean avalanche
size,$\ P(0)/P(1)$, diverges like $\exp({\mathrm{cst}}/(1/2-\alpha))$
 when $\alpha\rightarrow 1/2$.
These predictions
are compared in Fig.~4
to results obtained from the numerical solutions of 
Eq.~(\ref{eqa}) and (\ref{eqps}) for $K=2$ and different values of $\alpha$.
As shown in the insert, $1/P(0)$ vanishes linearly when $\alpha\rightarrow 1/2$
with a measured slope of 
$2.3$  close to the analytical prediction of $1/C=2$, the difference
between the two being quite compatible with higher order terms in $1/2-\alpha$.
The function $a(x)$ obtained by taking the inverse Fourier transform of
(\ref{af}) compares well 
to rescaled plots of $A(y)$
with $\eta(\alpha)$ chosen so as to make the different maxima coincide
\cite{no1}. Similar agreement is obtained between the analytic $\Pi(x)$ and
rescaled versions of $P(y)$ around $y=1/2$. Finally, the
extremely rapid growth of the mean avalanche size when
$\alpha\rightarrow 1/2$ agrees semi-quantitatively with the numerical
results shown in Fig.~2 but is itself an obstacle to a precise numerical
check of the predicted asymptotics. It makes it also difficult to avoid
finite size effects in numerical simulations when $\alpha\rightarrow
1/K$ given that the cut-off in the avalanche size distribution scales in
a mean field manner as the
square of the mean avalanche size \cite{lj}.

In conclusion, evidence has been presented that the random neighbor OFC
model has finite avalanches as soon as the model is non-conservative with
a mean avalanche size which increases extremely quickly as the
conservative limit is approached.
It would be interesting to assess the generality of this phenomenon and its
relevance for lattice models. 

\begin{figure}[htb]
\centerline{\epsfxsize=90mm \epsfysize=70mm\epsfbox{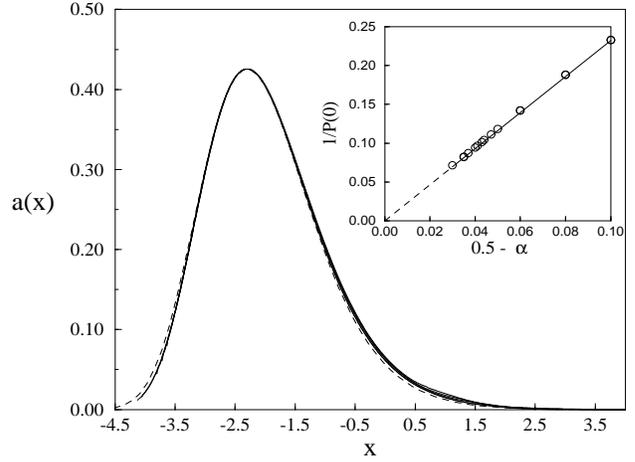}}
\caption
{$a(x)$(dashed line) and rescaled graphs $A(y) a^*/A^*$ {\em vs.}
$x=(y-1-2\eta(\alpha))\rho$ (lines) for $K=2$ and
$\alpha = 0.45, 0.456 $ and $0.463$. 
$a^*/A^*$ is the ratio of the curve maxima and
$\,\rho=A^*/(a^*\!\int\! A)$; 
$\rho/P(0)=1.145,1.117,1.089$  
for the graphs shown,
approaching
$1$ as expected when $\alpha\rightarrow 1/2$. 
Insert:
$1/P(0)$ versus $(1/2-\alpha)$; the straight line fit has a slope 
of 2.3}
\label{Gsim}
\end{figure}

\end{multicols}
\end{document}